\RequirePackage{amsmath}
\documentclass[epj]{svjour}

\usepackage{graphics}
\usepackage{graphicx}
\usepackage{color}
\usepackage[update,prepend]{epstopdf}
\usepackage{bbm}
\usepackage{amssymb} 
\newcommand{\nablas}{{\nabla\!}_s}
\newcommand{\laplaces}{{\nablas\!}^2}
\newcommand{\me}{\mathrm{e}}
\newcommand{\md}{\mathrm{d}}
\newcommand{\uu}[1]{\underline{\underline{#1}}}
\usepackage[dvipsnames]{xcolor}
\newcommand{\reA}[1]{\textcolor{black}{#1}}
\newcommand{\reB}[1]{\textcolor{black}{#1}}

\begin{document}

\titlerunning{Active Brownian motion of emulsion droplets}
\title{Active Brownian motion of emulsion droplets:\\
Coarsening dynamics at the interface and rotational diffusion}

\author{M. Schmitt \and H. Stark}

\offprints{}
\institute{                    
  Institut f{\"u}r Theoretische Physik, Technische Universit{\"a}t Berlin - Hardenbergstra{\ss}e 36, 10623 Berlin, Germany
}

\date{Received: date / Revised version: date}

\abstract{A micron-sized droplet of bromine water immersed 
in a surfactant-laden oil phase can swim\ \cite{thutupalli2011}. The bromine reacts with the surfactant at the droplet interface and generates a surfactant
mixture. It can spontaneously phase-separate due to solutocapillary Marangoni flow, which propels the droplet.
We model the system by a diffusion-advection-reaction equation for the mixture order parameter at the interface including
thermal noise and couple it to fluid flow. Going beyond previous work, we illustrate the coarsening dynamics of the surfactant mixture
towards phase separation in the axisymmetric swimming state. Coarsening proceeds in two steps: an initially slow growth of domain 
size followed by a nearly ballistic regime. On larger time scales thermal fluctuations in the local surfactant composition initiates random 
changes in the swimming direction and the droplet performs a persistent random walk, as observed in experiments. Numerical solutions
show that the rotational correlation time scales with the square of the inverse noise strength. We confirm this scaling by a perturbation 
theory for the fluctuations in the mixture order parameter and thereby identify the active emulsion droplet as an active
Brownian particle.}

\PACS{
{47.20.Dr}{Surface-tension-driven instability}\and
{47.55.D-}{Drops and bubbles}\and
{47.55.pf}{Marangoni convection}
     }

\maketitle

\section{Introduction}\label{sec:introduction}

In the past decade autonomous swimming of particles at low Reynolds number has attracted 
a tremendous amount of attention \cite{najafi2004,dreyfus2005,gauger2006,lauga2009,elgeti2015}.
Both, in the study of living organisms such as bacteria or algae or of artificial microswimmers
a plethora of exciting research subjects has evolved.
They include understanding the swimming mechanism \cite{fenchel2001,qiu2014,alizadehrad2015,maass2016}
and generic properties of microswimmers \cite{enculescu2011,romanczuk2012,zoettl2012,michelin2013},
their swimming trajectories \cite{howse2007,zaburdaev2011,theves2013,kummel2013},
and the study of their interaction with surfaces as well as obstacles \cite{volpe2011,drescher2011,majmudar2012,schaar2015}.
The study of emergent collective motion has opened up a new field in non-equilibrium statistical physics \cite{marchetti2013,ishikawa2008,evans2011,dunkel2013,alarcon2013,zoettl2014,hennes2014,pohl2014,zoettl2016}.

There are various methods to construct a microswimmer. One idea is to generate a slip velocity field close to the 
swimmer's surface using a phoretic mechanism.
A typical example of such an artificial swimmer is a micron-sized spherical Janus colloid,
which has an inherent polar symmetry. Its two faces are
made of different materials and thus differ in their physical or chemical properties \cite{walther2008}.
For example, a Janus particle with faces of different thermal conductivity moves if exposed to heat. 
The conversion of thermal energy to mechanical work
in a self-generated temperature gradient is called 
self-thermophoresis \cite{bickel2013}. 
Janus colloids also employ other phoretic mechanisms to become active \cite{golestanian2005,paxton2005,moran2011,buttinoni2012}.

A different realization of a self-propelled particle is an active emulsion droplet. The striking 
difference to an active Janus particle is the missing inherent polar symmetry.
Instead, the symmetry between front and back breaks spontaneously,
for example, in a subcritical bifurcation \cite{schmitt2013}.
The self-sustained motion of active droplets is due to a gradient in surface tension, which is usually caused by
an inhomogeneous density of surfactants. The resulting stresses set up a solutocapillary Marangoni flow directed
along the surface tension gradient that drags the droplet through the fluid.
An active droplet generates a flow field in the surrounding fluid typical for
the ``squirmer" \cite{lighthill1952,blake1971,downton2009,pak2014,schmitt2016}.
Originally, the squirmer was introduced to model the locomotion of microorganisms 
that propel themselves by a carpet of short active filaments 
called cilia beating in synchrony on their surfaces.
The squirmer flow field at the interface is then a coarse-grained model of the cilia carpet.

Active droplets have extensively
been studied  in experiments, including droplets in a bulk fluid \cite{hanczyc2007,toyota2009,thutupalli2011,kitahata2011,banno2012,ban2013,herminghaus2014,izri2014} 
and droplets on interfaces \cite{chen2009,bliznyuk2011}.
Theoretical and numerical studies address the drift bifurcation
of translational motion \cite{rednikov1994b,rednikov1994,velarde1996,velarde1998,yoshinaga2012}, 
deformable and contractile droplets \cite{tjhung2012,yoshinaga2014}, 
droplets in a chemically reacting fluid \cite{yabunaka2012}, 
droplets driven by nonlinear chemical kinetics \cite{furtado2008}, 
and the dif\-fu\-sion-\-ad\-vec\-tion-\-re\-ac\-tion equation for the dynamics of a surfactant mixture at 
the droplet interface \cite{schmitt2013}. A comprehensive review on active droplets is given in ref.\ \cite{maass2016}.

An active droplet, which swims due to solutocapillary Marangoni flow, 
has recently been realized \cite{thutupalli2011}. 
Water droplets with a diameter of $50-150\mu \mathrm{m}$
are placed into a surfactant-rich oil phase. 
The surfactants migrate to the droplet interface
where they form a dense monolayer.
Bromine dissolved in the water droplets
reacts with the surfactants at the interface. It saturates the double bond in the surfactant molecule
and the surfactant becomes weaker
than the original one. Hence, the ``bromination" reaction locally increases the interfacial surface tension. 
This induces Marangoni flow, which advects surfactants
and thereby further enhances the gradients in surface tension.
If the advective current exceeds the smoothing diffusion current, the surfactant mixture phase-separates. 
The droplet develops a polar symmetry and starts to move in a random direction,
which fluctuates around such that the droplet performs a persistent random walk.
While the droplet swims with a
typical swimming speed of $15\mu \mathrm{m/s}$, brominated surfactants are constantly replaced by non-bro\-mi\-na\-ted surfactants from the oil phase by means of desorption and adsorption. 
Finally, the swimming motion comes to an end 
when the fueling bromine is exhausted.

In ref.\ \cite{schmitt2013} we developed a diffusion-advection-reaction equation for the surfactant mixture at the 
droplet interface and coupled it to the axisymmetric flow field initiated by the Marangoni effect. In a parameter study we could then
map out a state diagram including the transition from the resting to the swimming state and an oscillating droplet motion.
In this paper we combine our theory with the full three-dimensional solution for the Marangoni flow, which we derived 
for an arbitrary surface tension field at the droplet interface in ref.\ \cite{schmitt2016}.
Omitting the constraint of axisymmetry and adding thermal noise to the dynamic equation of the surfactant mixture, we will
focus on two new aspects of droplet dynamics that we could not address in ref.\ \cite{schmitt2013}. 
First, while reaching the stationary uniaxial swimming state, the surfactant mixture phase-separates into the
two surfactant types. We illustrate the coarsening dynamics and demonstrate that it proceeds in two steps. An initially
slow growth of domain size is followed by a nearly ballistic regime. This is reminiscent to coarsening in the dynamic 
model H \cite{bray2003}.
Second, even in the stationary swimming state the surfactant composition fluctuates thermally and thereby
initiates random changes in the swimming direction, which diffuses on the unit sphere. As a result the droplet performs a 
persistent random walk, as observed in experiments \cite{thutupalli2011}, which we will characterize in detail.

The article is organized as follows. In sect.\ \ref{sec:model} we recapitulate our model of the active emulsion droplet from ref.\ \cite{schmitt2013} and generalize it to a droplet without the constraint of axisymmetry.
While sect.\ \ref{sec:finite_volume_method} explains
the numerical method to solve the diffusion-advection-reaction equation on the droplet surface,
the following two sections contain the results of this article.
Section\ \ref{sec:dynamics_towards_the_swimming_state} 
describes the coarsening dynamics of the surfactant mixture before reaching the steady swimming state
and sect.\ \ref{sec:dynamics_of_the_swimming_state} characterizes the persistent random walk of the droplet in the
swimming state. The article concludes in sect.\ \ref{sec:conclusions}.

\section{Model of an active droplet}\label{sec:model}

In order to model the dynamics of the active droplet,
we follow our earlier work \cite{schmitt2013}. We use
a dynamic equation for the surfactant mixture 
at the droplet interface that includes all the relevant processes. We assume that the surfactant completely covers the droplet interface
without any intervening solvent. We also assume that the head area of both types of surfactant molecules (brominated and non-brominated) is the same. Denoting the brominated surfactant density by $c_1$ and the non-brominated density by $c_2$, we can therefore set $c_1+c_2=1$. We then take the concentration difference between brominated and non-brominated surfactants as an order parameter $\phi=c_1-c_2$.
In other words $\phi=1$ corresponds to fully brominated and $\phi=-1$ to fully non-brominated
surfactants and $c_1=(1+\phi)/2$ and $c_2=(1-\phi)/2$. Finally, we choose a constant droplet radius $R$.

\subsection{Diffusion-advection-reaction equation}\label{subsec:diffusion-advection-reaction_equation}

The dynamics of the order parameter $\phi$ at the droplet interface can be expressed as \cite{schmitt2013}:
\begin{equation}
   \partial_t \phi=-\nablas\cdot(\mathbf{j}_D+\mathbf{j}_A) - \tau_R^{-1} (\phi-\phi_{\mathrm{eq}})+\zeta(\mathbf{r},t)\;,\label{eq:conti}
\end{equation}
which we formulate in the form of a continuity equation with 
an additional source and thermal noise ($\zeta$) term.
$\nablas=(\mathbf{1}-\mathbf{n}\otimes\mathbf{n})\nabla$ 
stands for the directional gradient on a sphere with radius $R$,
where $\nabla$ is the nabla operator and $\mathbf{n}$ the surface normal. The 
current is split up into a diffusive part $\mathbf{j}_D$ and an advective part $\mathbf{j}_A$, which arises due to the Marangoni effect. 
We summarize them below and in sect.\ \ref{subsec:marangoni_flow}.
The source term describes the bromination reaction as well as desorption of brominated and adsorption of non-brominated surfactants to and from the outer fluid.  Both processes tend to establish an equilibrium mixture with order parameter $\phi_{\mathrm{eq}}$
during the characteristic relaxation time $\tau_R$.
Ad- and desorption dominate for $\phi_{\mathrm{eq}}<0$ 
while bromination dominates for $\phi_{\mathrm{eq}}>0$. 
The source term is a simplified phenomenological description for the ad- and desorption of surfactants. 
A more detailed model would include fluxes from and to the bulk fluid \cite{ipac2002}.
We will explain the thermal noise term further below.

The general mechanism 
of eq.\ (\ref{eq:conti}) to initiate steady Marangoni flow
is as follows. The diffusive current $\mathbf{j}_D$ smoothes out gradients in $\phi$, while the advective Marangoni current $\mathbf{j}_A$ amplifies gradients in $\phi$. Hence, $\mathbf{j}_D$ and $\mathbf{j}_A$ are competing and as soon as $\mathbf{j}_A$ dominates over $\mathbf{j}_D$, $\phi$ experiences phase separation. As a result, the resting state becomes unstable and the droplet starts to swim. 

We now summarize features of the diffusive current $\mathbf{j}_D$, more details can be found in ref.\ \cite{schmitt2013}.
We formulate a Flory-Huggins free energy density in terms of the order parameter of the surfactant mixture,
which includes entropic terms and interactions between the different types of surfactants:
\begin{equation*}
\left. \begin{array}{ll}
f(\phi)=\frac{k_B T}{\ell^2} \left[ \frac{1+\phi} 2 \ln \frac{1+\phi} 2+\frac{1-\phi} 2 \ln \frac{1-\phi} 2\right.\\
\\
\left.-\frac 1 4 (b_1+b_2+b_{12})- \frac {\phi} 2 (b_1-b_2)-\frac{\phi^2} 4 (b_1+b_2-b_{12}) \right] ,\end{array} \right. \label{eq:free_energy}
\end{equation*}
Here, $\ell^2$ is the head area 
of a surfactant at the interface. We introduce dimensionless parameters $b_1$ ($b_2$) to characterize the interaction between brominated (non-brominated) surfactants and $b_{12}$ describes the interaction between the two types of surfactants.
The diffusive current is now driven by a gradient in the chemical potential derived from the total free energy 
functional $F[\phi]=\iint f(\phi)\, \md A$:
\begin{equation}
   \mathbf{j}_D=-\lambda\nablas \frac{\delta F}{\delta \phi}=-D \left[\frac 1 {1-\phi^2}-\frac 1 2 (b_1+b_2-b_{12})\right]\nablas\phi\;, \label{eq:j_d}
\end{equation}
where the Einstein relation $D=\lambda k_B T/\ell^2$ relates the interfacial diffusion constant $D$ to the mobility $\lambda$. 
To rule out a double well form of $f(\phi)$, which would generate phase separation already in thermal equilibrium,
we only consider $b_1 +b_2 -b_{12} < 2$. This also means that the diffusive current $\mathbf{j}_D\propto-\nablas\phi$
is for all $\phi$ indeed directed against $\nablas\phi$. In the following we assume $b_{12}=(b_1+b_2)/2$ and therefore
require $b_1+b_2<4$.

We formulate the thermal noise term in eq.\ (\ref{eq:conti}) as Gaussian white noise with zero mean following 
ref.\ \cite{desai2009}:
\begin{subequations}\label{eq:fluc_diss}
\begin{eqnarray}
\langle\zeta\rangle&=&0\;,\\
\langle\zeta(\mathbf{r},t)\zeta(\mathbf{r}',t')\rangle&=&-2k_BT\lambda\laplaces\delta(\mathbf{r}-\mathbf{r}')\delta(t-t')\;.\label{eq:fluc_diss_b}
\end{eqnarray}
\end{subequations} 
Here, the strength of the noise correlations is connected to the mobility $\lambda$ of the diffusive current via the 
fluctuation-dissipation theorem.
In order to close eq.\ (\ref{eq:conti}), we now discuss the advective Marangoni current $\mathbf{j}_A$.

\subsection{Marangoni flow}\label{subsec:marangoni_flow}

The advective current for the order parameter $\phi$ is given by
\begin{equation}
 \mathbf{j}_A=\phi \mathbf{u}|_R\;,\label{eq:j_a}
\end{equation}
where $\mathbf{u}|_R$ is the flow field at the droplet interface.
It is driven by a non-uniform surface tension $\sigma$
and therefore called Marangoni flow \cite{chandrasekhar1961,ipac2002}.
In our case, we have a non-zero surface divergence $\nablas\cdot\mathbf{u}|_R \ne 0$. In fact, it can be shown that an 
incompressible surface flow cannot lead to propulsion of microswimmers \cite{stone1996}.

In order to evaluate $\mathbf{u}|_R$, one has to solve the Stokes equation for the flow field $\mathbf{u}(\mathbf{r})$ surrounding the spherical droplet ($r>R$) as well as for the flow field $\hat{\mathbf{u}}(\mathbf{r})$ 
inside the droplet ($r<R$).
Both solutions are matched at the droplet interface by the condition \cite{ipac2002},
\begin{equation}
\nablas \sigma = \mathbf{P}_s \left. (\mathbf{T}-\hat{\mathbf{T}}) \mathbf{e}_r\right|_{r=R} \; ,
\label{eq.boundary}
\end{equation}
where $\mathbf{P}_s =  \mathbf{1} - \mathbf{e}_r \otimes \mathbf{e}_r$ is the surface projector.
Equation\ (\ref{eq.boundary}) means that a gradient in surface tension $\sigma$ is compensated by a jump in viscous shear stress.
Here, $\mathbf{T}=\eta[\nabla\otimes \mathbf{u}+(\nabla\otimes \mathbf{u})^T]$ is the viscous shear stress tensor of
a Newtonian fluid with viscosity $\eta$ outside of the droplet and the same relation holds for $\hat{\mathbf{T}}$ 
of the fluid with viscosity $\hat\eta$ inside the droplet. We have performed this evaluation in ref.\ \cite{schmitt2016}
for a given surface tension field and only summarize here the results relevant for the following. Alternative derivations
are found in ref.\ \cite{blawzdziewicz2000,hanna2010,schwalbe2011,pak2014b}.

In spherical coordinates the Marangoni flow field $\mathbf{u}|_R$
at the interface reads \cite{schmitt2016,blawzdziewicz2000,hanna2010,schwalbe2011}
\begin{equation}
\mathbf{u}|_R=\frac{-\eta}{2({\eta}+\hat{\eta})}\mathbf{v}_D+\frac{1}{{\eta}+\hat{\eta}}\sum_{l=1}^{\infty}\limits\sum_{m=-l}^l\limits\frac{R\ s_l^m}{2l+1}\nablas Y_l^m \;,\label{eq:u_solution}
\end{equation}
with spherical harmonics $Y_l^m(\theta,\varphi)$ given in appendix \ref{sec:spherical_harmonics}. 
Here, 
\begin{equation}
 s_l^m=\iint \sigma(\theta,\varphi) \overline{Y}_l^m(\theta,\varphi)\ \md\Omega  \label{eq:s_l^m}
\end{equation}
are the expansion coefficients of the surface tension, where $\overline{Y}_l^m$ means complex conjugate of
$Y_l^m$, and \cite{schmitt2016,hanna2010,pak2014b}
\begin{equation}
\mathbf{v}_D = v_D \mathbf{e}
= \frac 1 {\sqrt{6\pi}}\frac{1}{2\eta+3\hat{\eta}}
   \left(\begin{array}{c} 
        s_1^1-s_1^{-1} \\  
        i\left(s_1^1+s_1^{-1}\right)  \\ 
        -\sqrt 2 s_1^0 
\end{array}\right)\;.
\label{eq:velocity_vector}
\end{equation}
is the droplet velocity vector. It is solely given by the dipolar coefficients ($l=1$) of the surface tension and
determines propulsion speed $v_D \geq 0$ as well as the swimming direction $\mathbf{e}$ with $|\mathbf{e}|=1$.
Note that by setting $m=0$, eqs.\ (\ref{eq:u_solution})-(\ref{eq:velocity_vector}) reduce to the case of an axisymmetric 
droplet swimming along the $z$-direction, as studied in ref.\ \cite{schmitt2013}.

In ref. \cite{schmitt2016} we give
several examples of flow fields $\mathbf{u}|_R$. In general, Marangoni flow is directed along gradients in surface tension, 
\emph{i.e.} $\mathbf{u}|_R\parallel\nablas\sigma$.
This is confirmed by eq. (\ref{eq:u_solution}) 
and also clear from fig.\ \ref{fig:2} (b), which we discuss later.
{However, according to eq.\ (\ref{eq:u_solution}) higher modes of surface tension contribute with a decreasing
coefficient \cite{schmitt2016}. 
Note the velocity field in eq.\ (\ref{eq:u_solution}) is given in a frame of reference that 
moves with the droplet's center of mass but the directions of its axis are fixed in space and do not rotate with the droplet.
Finally, the velocity fields inside ($\hat{\mathbf{u}}$) and outside ($\mathbf{u}$) of the droplet in both the droplet and 
the lab frame can be found in the appendix of ref.\ \cite{schmitt2016}.

The surface tension necessary to calculate $\mathbf{v}_D$ and $\mathbf{u}|_R$
is connected to the order parameter $\phi$ by
the equation of state, $\sigma=f-\frac {\partial f}{\partial c_{1}} c_1-\frac {\partial f}{\partial c_{2}}c_2$,  
which gives \cite{schmitt2013}
\begin{equation}
 \sigma(\phi)=\frac{k_B T }{\ell^2}(b_1-b_2)\left( \frac 3 8 \frac{b_1+b_2}{b_1-b_2} + \frac 1 2 \phi+\frac 1 8 \frac{b_1+b_2}{b_1-b_2} \phi^2 \right) \;.\label{eq:sigma}
\end{equation}
This implies that for $b_1>b_2>0$, $\nablas\phi$ points along $\nablas\sigma$. Moreover, 
since the Marangoni flow $\mathbf{u}|_R$ is oriented along $\nablas\sigma$, 
as noted above, we conclude that for $\phi>0$
the advective current $\mathbf{j}_A = \phi \mathbf{u}|_R$ 
points ``uphill", \emph{i.e.}, in the direction of $\nablas\phi$, in contrast to $\mathbf{j}_D$ \cite{schmitt2013}. 

This completes the derivation of the surface flow field $\mathbf{u}|_R$  
as a function of the expansion coefficients $s_l^m$
of the surface tension.
Together with the equation of state 
$\sigma(\phi)$ the advective current $\mathbf{j}_A$ in eq.\ (\ref{eq:j_a}) is specified. Finally, using the diffusion current 
$\mathbf{j}_D$ from eq.\ (\ref{eq:j_d}), the diffusion-advection-reaction equation (\ref{eq:conti}) becomes a closed 
equation in $\phi$.

The swimming emulsion droplet is an example of a spherical microswimmer, a so-called squirmer \cite{lighthill1952,blake1971,downton2009,pak2014,schmitt2016}. 
Squirmers are often classified by means of the so-called squirmer parameter $\beta$ \cite{lauga2009}. 
When $\beta<0$, the surface flow dominates at the back of the squirmer, similar to the flow field of the bacterium \emph{E. coli}. 
Since such a swimmer pushes fluid 
outward along its major axis,
it is called a 'pusher'. Accordingly, a swimmer with $\beta>0$ is called a 'puller'. 
The algae \emph{Chlamydomonas} is a biological example of a puller.
Swimmers with $\beta=0$ are called 'neutral'.

For an axisymmetric emulsion droplet swimming along the $z$-direction, the squirmer parameter is given by
\begin{equation}
 \beta=-\sqrt{\frac{27}5}\frac{s_2^0}{|s_1^0|}\;,\label{eq:squirmer_parameter}
\end{equation}
with coefficients $s_l^m$ from the multipole expansion (\ref{eq:s_l^m}) of the surface tension $\sigma$ \cite{schmitt2016}. 
A generalization of this formula to 
droplets without axisymmetry and
swimming in arbitrary directions
is derived in ref.\ \cite{schmitt2016}. 
The relevant expressions are presented in
appendix \ref{sec:beta_details}.

\subsection{\mbox{Reduced dynamic equations and system parameters}}\label{subsec:system_parameters}

In order to write eq.\ (\ref{eq:conti}) in reduced units,
we rescale time by the characteristic
diffusion time $\tau_D=R^2/D$ and lengths by droplet radius $R$,
and arrive at
\begin{equation}
   \partial_{t}\phi=-\nablas\cdot({\mathbf{j}}_D+M\phi{\mathbf{u}}|_R) - \kappa (\phi-\phi_{\mathrm{eq}})+\xi{\zeta}(\mathbf{r},t)\;,\label{eq:conti_dimless}
\end{equation}
where the Gaussian noise variable fulfills 
\begin{equation}
 \langle\zeta(\mathbf{r},t)\zeta(\mathbf{r}',t')\rangle=-2 \laplaces\delta(\mathbf{r}-\mathbf{r}')\delta(t-t')\;.\label{eq:fluc_diss_b_dimless}
\end{equation}
The dimensionless velocity field at the interface and the droplet velocity vector read, respectively,
\begin{subequations}\label{eq:u_solution_and_velovec_dimless}
\begin{eqnarray}
{\mathbf{u}}|_R&=&-\frac{\mathbf{v}_D}{2}+\sum_{l=1}^{\infty}\limits\sum_{m=-l}^l\limits\frac{{s}_l^m}{2l+1} \nablas Y_l^m\;, \label{eq:u_solution_dimless}\\
 \mathbf{v}_D &=& v_D \mathbf{e}
= \frac 1 {\sqrt{6\pi}(2+3\nu)}
   \left(\begin{array}{c} 
        s_1^1-s_1^{-1} \\  
        i\left(s_1^1+s_1^{-1}\right)  \\ 
        -\sqrt 2 s_1^0 
\end{array}\right)\;.
\label{eq:dimless_velocity_vector}
\end{eqnarray}
\end{subequations}}
All quantities in eqs.\ (\ref{eq:conti_dimless}) 
and (\ref{eq:u_solution_and_velovec_dimless}), including $\mathbf{j}_D$, $\mathbf{u}|_R$, $t$, $\nablas$, $\zeta$, and $\mathbf{v}_D$, are 
from now on dimensionless,
although we use the same symbols as before. Writing the dynamics equations in reduced units, introduces the relevant
system parameters $M, \nu, \kappa, \phi_{\mathrm{eq}}$, and $\xi$,
which we discuss now.

The Marangoni number $M$ quantifies the strength of
the advective current in eq. (\ref{eq:conti_dimless}) 
and is given by $M=\frac{(b_1-b_2)R}{\lambda(\eta+\hat\eta)}$. 
It is the most important parameter of our model, as it determines whether the droplet swims.
In eq.\ (\ref{eq:u_solution_dimless}) we introduced the ratio of 
shear viscosities, $\nu=\hat\eta / \eta$, for the fluids inside and outside of the droplet, respectively.
In our study we consider a water droplet suspended in oil and set $\nu\approx 1/36$ \cite{thutupalli2011}. 
The interaction parameters $b_1$ and $b_2$ not only
appear in $M$ but also as $b_1+b_2$ in the diffusive current in eq.\ (\ref{eq:j_d}) and in the equation of state $\sigma(\phi)$ in eq.\ (\ref{eq:sigma}). 
Therefore, they need to be set individually. 
Assuming the head area of a surfactant $\ell^2$ to be on the order of $\mathrm{nm}^2$, we can fit eq.\ (\ref{eq:sigma}) to
the experimental values $\sigma(\phi=1)\approx 2.7 \mathrm{mN}/\mathrm{m}$ and $\sigma(\phi=-1)\approx 1.3 \mathrm{mN}/\mathrm{m}$\cite{thutupalli2011} to find  $b_1\approx 0.6$ and $b_2\approx 0.3$.
We keep these values fixed throughout the article.

Parameter $\kappa=\tau_D/\tau_R$ tunes the ratio between diffusion and 
relaxation time and
the equilibrium order parameter $\phi_{\mathrm{eq}}$ measures whether ad- and desorption
of surfactants
($\phi_{\mathrm{eq}}<0$) or bromination ($\phi_{\mathrm{eq}}>0$) dominates. In this study we set $\kappa=0.1$ and $\phi_{\mathrm{eq}}=0.5$. A parameter study for these parameters can be found in \cite{schmitt2013}.
Finally, the reduced noise strength $\xi=\ell/R \propto 1/\sqrt{N}$,
where $N$ is the total number of surfactants at the droplet interface, 
connects the the droplet size $R$ to the molecular length scale $\ell$.

\reA{The following sect. \ref{sec:finite_volume_method} describes, how we solved the dynamic equation (\ref{eq:conti_dimless}) numerically. Readers not interested in the details can proceed immediately to sect. \ref{sec:dynamics_towards_the_swimming_state}, where we present our first results.}

\section{Finite volume method on a sphere} \label{sec:finite_volume_method}

To numerically solve the rescaled dynamic equation (\ref{eq:conti_dimless})  for the order parameter field $\phi$, we 
had to decide on an appropriate method.
The most widely used numerical methods for solving partial differential equations 
are the finite difference method (FDM), the finite element
method (FEM), and the finite volume method (FVM) \cite{ferziger1996,eymard2000}. We ruled 
out FDM due to numerical complications of its algorithm
with spherical coordinates. They are most 
appropriate for the spherical droplet surface but one needs to define
an axis within the droplet.
The FEM is also very delicate when writing
a numerically stable code for our model. This is mainly due to the advective term in eq.\ (\ref{eq:conti_dimless}), 
which commonly causes difficulties in FEM routines \cite{ferziger1996}.
In contrast, the FVM is especially suited for solving 
continuity equations. Therefore, it is
much more robust for field equations that incorporate advection 
and we chose it for solving eq.\ (\ref{eq:conti_dimless}) on the droplet surface.

In order to generate a two-dimensional
FVM mesh that is as uniform as possible and quasi-isotropic on a sphere, we chose a geodesic grid based on a refined 
icosahedron \cite{baumgardner1985}. An icosahedron has $f_0=20$ equilateral 
triangles as faces and $v_0=12$ vertices. 
In each refinement step, each triangle is partitioned into 
four equilateral triangles and the three
new vertices are projected onto the 
unit sphere enclosing the icosahedron.
Hence, after the $n$-th refinement step,
the resulting mesh has $f_n=4^n f_0$ triangular faces and $v_n=v_{n-1}+\frac 3 8 4^n f_0$ grid 
points.
\footnote{Each face has three edges
and every edge belongs to two faces, hence the number of edges is 
$e_n=\frac 3 2 f_n$. In a refinement
step one new grid point is placed on the middle of each edge and
$v_n=v_{n-1}+e_{n-1}$. Thus,
$v_n=v_{n-1}+\frac 3 8 4^n f_0$, with $v_1=42, v_2=162, v_3=642, v_4=2562$.}
The ``finite volume'' then refers to a small volume (in this case 
an area) surrounding each grid point of the mesh. 
Thus we have to construct the Voronoi diagram of the triangular mesh. 
The Voronoi diagram consists of $v_n$ elements, $12$ of which are pentagons
associated with the vertices of the original icosahedron
while the rest are hexagons.
Unless otherwise noted we use a 
Voronoi mesh with $v_3=642$ 
FVM elements. The geodesic icosahedral grid is a standard grid in geophysical fluid dynamics. A comprehensive review 
on numerical methods in geophysical fluid dynamics can be found in \cite{pudykiewicz2006}.

\begin{figure}
\centering
\includegraphics[width=0.4\textwidth, angle=0, trim=0mm 0mm 0mm 0mm, clip]{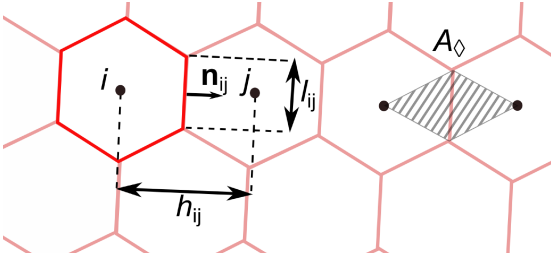}
\caption{Finite volume element $i$ with neighboring element $j$. 
The relevant lengths and normal vector are sketched.
}
\label{fig:1}
\end{figure}

In the following we will outline how we convert the diffusion-advection-reaction equation (\ref{eq:conti_dimless}) to a set of 
ordinary differential equations for a vector $\underline{\phi}$ comprising the values $\phi_i$ of the order parameter field
at the center points of all FVM elements.
FVM was developed for treating current densities in a continuity equation and we illustrate the procedure for the diffusion term of
eq.\ (\ref{eq:conti_dimless}).
We start by integrating over element $i$ with area $A_i$ and use the divergence theorem, where $\mathbf{n}_i$ is the 
outward normal at the element boundary:
\begin{subequations}\label{eq:fvm}
\begin{eqnarray}
\iint\limits_{A_i} \nablas\cdot \mathbf{j}_D\, \md A&=&\int_{\partial A_i}\limits \mathbf{j}_D \cdot \mathbf{n}_i\, \md S
= \sum_{j=1}^{N} \mathbf{j}_D \cdot \mathbf{n}_{ij} l_{ij} 
\qquad \label{eq:fvm1}\\
&=& -\sum_{j=1}^{N}  
D(\phi_i,\phi_j)
\frac{\phi_j-\phi_i}{h_{ij}}  l_{ij}  = \uu{D}^{i} \underline{\phi}\,.  \; \qquad
\label{eq:fvm2}
\end{eqnarray}
\end{subequations}
In the last term of eq.\ (\ref{eq:fvm1}), the line integral is 
converted into a sum over the $N$ straight element boundaries of length $l_{ij}$
and $\mathbf{n}_{ij}$ is the normal vector at the corresponding boundary.
Figure\ \ref{fig:1} illustrates the relevant quantities.
In the second line the directional derivative $\mathbf{n}_{ij} \cdot \nablas\phi$
resulting from $\mathbf{j}_D$ in eq.\ (\ref{eq:j_d}) is approximated by a difference quotient. 
The prefactor in $\mathbf{j}_D$, which we abbreviated by $D(\phi_i,\phi_j)$ in eq.\ (\ref{eq:fvm2}), also contains $\phi$.
It is interpolated at the boundary between elements $i$ and $j$ 
by means of the
central differencing scheme as $(\phi_i + \phi_j) /2$.
Finally, we write the whole term as the product of local
diffusion matrix $\uu{D}^i$ and vector $\underline{\phi}$.
After applying this technique to all elements, the matrices $\uu{D}^i$ are 
combined into one matrix $\uu{D}$ for the whole mesh. 

The same procedure is carried out for the advective term in eq.\ (\ref{eq:conti_dimless}) but discretizing $\mathbf{j}_A=M\phi\mathbf{u}|_R$ needs more care. While $\mathbf{u}|_R$ is directly calculated at the boundary between elements $i$ and $j$, the order parameter $\phi$ is treated differently. 
If the local Peclet number $\mathrm{Pe}={h_{ij}\, M |\mathbf{u}|_R|}/{D(\phi_i,\phi_j)}$ 
is larger than $2$, the central differencing scheme fails to converge. Instead
a so-called upwind scheme is used,
which takes into account the direction of 
flow \cite{ferziger1996}. 
For outward oriented flow, \emph{i.e.} $\mathbf{u}|_R\cdot\mathbf{n}_{ij}>0$, one uses the element order parameter 
$\phi_i$, while for inward flow, \emph{i.e.} $\mathbf{u}|_R\cdot\mathbf{n}_{ij}<0$, one uses the order parameter of the 
neighboring element $\phi_j$. In the case $\mathrm{Pe}<2$, $\phi$ is interpolated
by the central difference
$(\phi_i + \phi_j) /2$.

Finally, the linear terms in $\phi$ and its time derivative
are simply approximated by $\phi_i$ and $\dot\phi_i$.
In the end, we are able to write the discretized eq. (\ref{eq:conti_dimless}) as a matrix equation for the vector $\underline{\phi}$:
\begin{equation}
\uu{M} \,\underline{\dot\phi}=\uu{D} \,\underline{\phi}-M \uu{A}\,\underline{\phi}-\kappa \uu{M}\left(\underline{\phi}-\underline{\phi}_{\,\mathrm{eq}}\right)+ 2 \cdot 12^{1/4} \xi\underline{z} \; ,
\label{eq:discretize}
\end{equation}
where the diagonal matrix $\uu{M}$ carries the areas of the elements, 
and with diffusion matrix $\uu{D}$, advection matrix $\uu{A}$, and element noise vector $\underline{z}$, 
which describes typical Gaussian white noise with zero mean and variance one,
\begin{subequations}
\begin{eqnarray}
\langle\underline{z}(t)\rangle&=&\underline{0}\;,\\
\langle\underline{z}(t)\otimes\underline{z}(t')\rangle&=&\uu{1}\delta(t-t')\;.\label{eq:fluc_diss_element}
\end{eqnarray}
\end{subequations}
In appendix \ref{sec:element_moise_vector} we derive eq.\ (\ref{eq:fluc_diss_element}) by 
integrating eq.\ (\ref{eq:fluc_diss_b_dimless}) over two FVM elements $i$ and $j$.
Finally, the set of stochastic differential equations are integrated in time by a standard Runge-Kutta scheme. 

In the following we present results obtained with the described numerical scheme.

\section{Dynamics towards the swimming state}\label{sec:dynamics_towards_the_swimming_state}

This section focuses on the dynamics of the active emulsion droplet from an initial resting state with swimming speed $v_D=0$ to a stable swimming state with swimming speed $v_D>0$. After a comparison with 
the axisymmetric model of the droplet from our previous work \cite{schmitt2013}, where we also did not include thermal fluctuations, we investigate the coarsening dynamics of the order parameter $\phi$ at the droplet interface
while reaching the swimming state. 

\begin{figure}
\includegraphics[width=0.7\textwidth, angle=-90, trim=0mm 0mm 0mm 140mm, clip]{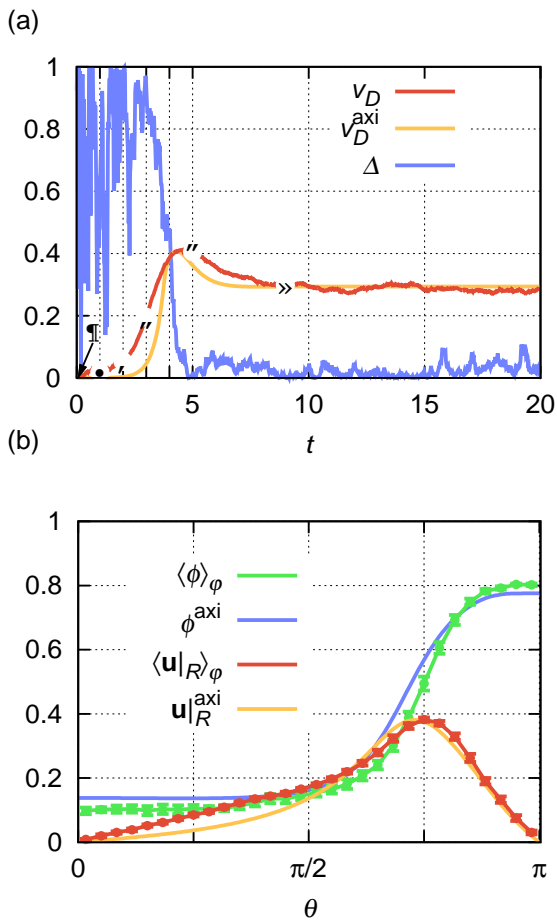}
\caption{(a) Droplet swimming speed $v_D(t)$ of an active droplet from a simulation with $v_4=2562$ FVM elements. 
Order parameter profiles $\phi$ at the time steps marked with numbers are given in fig.\ \ref{fig:4}(a). 
For comparison, we plot $v_{D}^{\mathrm{axi}}$ of the axisymmetric model taken from fig.\ 2 (a) of ref.\ \cite{schmitt2013}
but on a different scale.
We also show biaxiality parameter $\Delta$ of the order parameter field
defined in eq.\ (\ref{eq:biaxiality}).
\reA{Noise strength is set to $\xi=10^{-3}$, Marangoni number to $M=3$, reduced
reaction rate to $\kappa=0.1$, and equilibrium order parameter value to $\phi_{\mathrm{eq}}=0.5$.}
(b) Order parameter profile $\langle\phi \rangle_{\varphi}$
and velocity field $\langle \mathbf{u}|_R \rangle_{\varphi}$
at $t=20$, averaged about the swimming axis $\mathbf{e}$
as indicated by $\langle\dots\rangle_{\varphi}$ and defined in appendix\ \ref{sec:averages}.
The front of the droplet corresponds to the polar angle $\theta=0$.
For comparison, we plot $\phi^{\mathrm{axi}}$ and $\mathbf{u}|_R^{\mathrm{axi}}$ from the axisymmetric model 
taken from fig.\ 1 of ref.\ \cite{schmitt2013}.
Note that the Marangoni flow $\mathbf{u}|_R$ is directed along the gradients of $\phi$ and surface tension $\sigma$.
}\label{fig:2}
\end{figure}

\subsection{Swimming speed $v_D$}\label{subsec:swimming_speed}

In order to test the simulation method, we start our analysis with a set of parameters, for which we found a swimming state in the inherent axisymmetric model \cite{schmitt2013}. 
They are given by Marangoni number $M=3$, reduced
reaction rate $\kappa=0.1$, and equilibrium order parameter value $\phi_{\mathrm{eq}}=0.5$. We keep these values fixed 
throughout the following unless otherwise noted. The initial condition 
for solving eq.\ (\ref{eq:discretize}) is an order parameter field that fluctuates around $\phi_{\mathrm{eq}}$:
$\phi(\theta,\varphi)=\phi_{\mathrm{eq}} + \delta\phi(\theta,\varphi)$.
The small fluctuations $\delta\phi(\theta,\varphi) \ll 1$ are realized by random numbers drawn from the normal distribution $\mathcal{N}(\phi_{\mathrm{eq}},\alpha^2)$
with mean $\phi_{\mathrm{eq}}$ and variance $\alpha^2=10^{-5}$
and added at the grid points of the simulation mesh. 
Furthermore we set the noise strength to $\xi=10^{-3}$. Figure\ \ref{fig:2}(a) shows the droplet swimming 
speed $v_D$ as a function of elapsed time.

\begin{figure}
\includegraphics[width=0.28\textwidth, angle=-90, trim=0mm 0mm 0mm 0mm, clip]{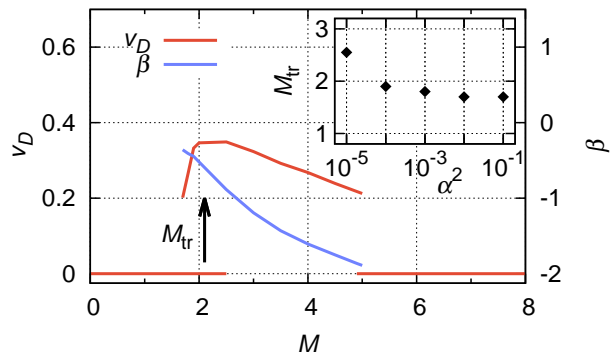}
\caption{Droplet swimming speed $v_D$ and squirmer parameter $\beta$ plotted versus Marangoni number $M$ 
for zero thermal noise $\xi=0$. At the transition Marangoni number $M_{\mathrm{tr}}$, $v_D$ jumps to a non-zero value 
indicating a subcritical bifurcation.
Inset: $M_{\mathrm{tr}}$ versus noise strength $\alpha^2$, with which the initially uniform order parameter
profile is disturbed.
The swimming regime terminates at an upper bifurcation, see also ref.\ \cite{schmitt2013}.
}\label{fig:3}
\end{figure}

First of all, we notice the good agreement with the corresponding graph of 
$v_{D}^{\mathrm{axi}}$ of the axisymmetric system of ref.\ \cite{schmitt2013}, which we also 
plot in fig.\ \ref{fig:2}(a). 
The same applies to the order parameter profile $\phi$ and the 
surface velocity field $\mathbf{u}|_R$ of the swimming state, when averaged about the swimming axis 
$\mathbf{e}$, see fig.\ \ref{fig:2}(b).
Thus, the full three-dimensional
description presented in this work is consistent with the axisymmetric model of ref.\ \cite{schmitt2013}. 
The same is true for the squirmer parameter $\beta$ from eq.\ (\ref{eq:beta_full}), for which we find $\beta\approx -1.2$ for $M=3$. This is fairly close to the value of the axisymmetric model
($\beta\approx -0.8$) and confirms that the swimming active droplet is a pusher.

We stress that the Marangoni number $M$ is the crucial parameter in our model, as it determines whether the droplet rests or swims. 
For small $M$, the homogeneous state $\phi=\phi_{\mathrm{eq}}$ is stable, 
\emph{i.e.}, any disturbance $\delta\phi$ of the 
initially uniform $\phi$ is damped by the diffusion and reaction terms of eq.\ (\ref{eq:conti_dimless}). As a result, the droplet rests.
The transition to the swimming state occurs at increasing Marangoni number $M$ via a subcritial bifurcation as
illustrated in fig.\ \ref{fig:3}, which shows swimming speed $v_D$ and squirmer parameter $\beta$ plotted versus $M$.
We use here a system without thermal noise, \emph{i.e.}, $\xi=0$, in order to monitor the complete transition region of
the subcritical bifurcation.
At a transition value $M_{\mathrm{tr}}$ the advective term of eq. (\ref{eq:conti_dimless}) overcomes the damping terms. 
The homogeneous state becomes unstable and the droplet starts to swim with a finite swimming speed $v_D$. 
As usual for a subcritical bifurcation, the transition to the swimming state takes place in a finite interval of $M$.
There, the transition Marangoni number $M_{\mathrm{tr}}$ depends on the initial
disturbance strength $\alpha^2$ of the uniform order parameter profile.
The inset of fig.\ \ref{fig:3} confirms this statement.
Next, we will discuss the biaxial evolution and the
coarsening dynamics of the order parameter field, which we could not study in the
axisymmetric description.

\subsection{Transient biaxial dynamics}\label{subsec:transient_biaxial_dynamics}

\begin{figure*}
\includegraphics[width=0.53\textwidth, angle=-90, trim=20mm 0mm 20mm 0mm, clip]{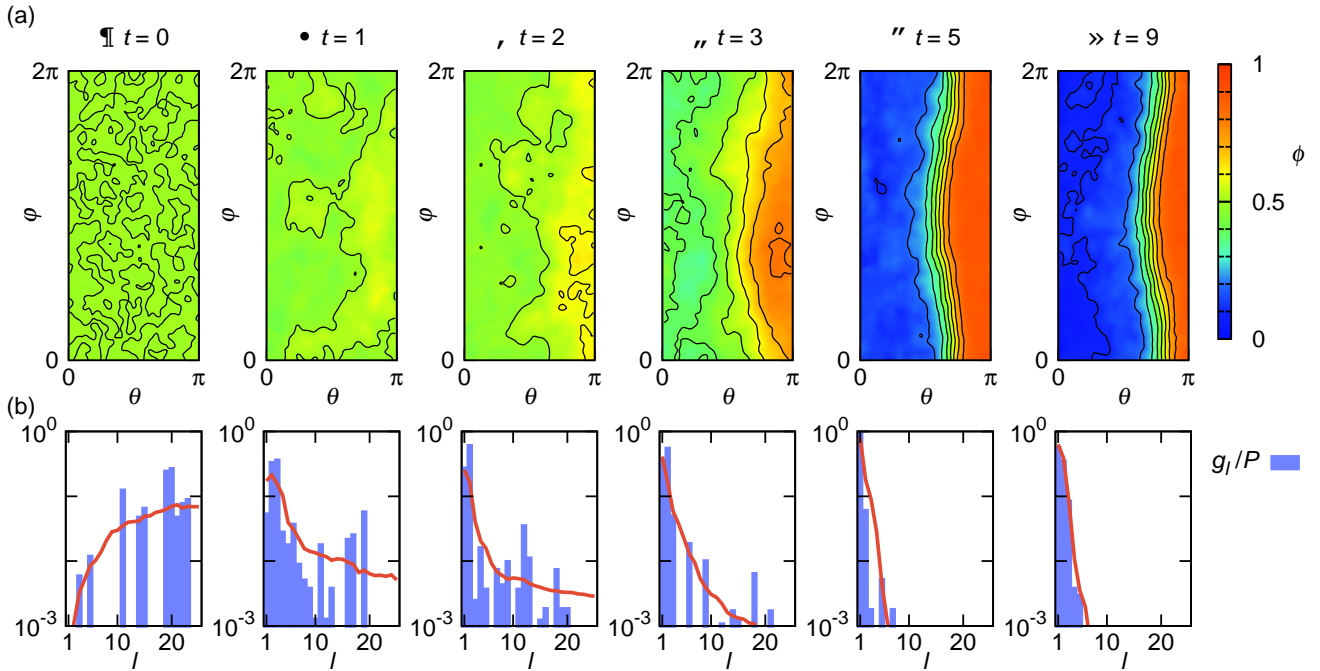}
\caption{(a)
Color-coded order parameter profile $\phi(\theta,\varphi)$ at various time steps
in the coordinate frame of the droplet, where the front of the droplet is located at $\theta=0$. 
Lines of equal $\phi$ are drawn. The time snapshots are indicated in fig.\ \ref{fig:2} (a) in
the curve for $v_D$ (same simulation run).
The relevant parameters are: $M=3$, $\kappa = 0.1$, $\phi_{\mathrm{eq}} = 0.5$, and $\xi=10^{-3}$.
(b) The bar charts show the normalized polar power spectrum $g_l /P$ of surface tension $\sigma$ for the profiles in (a). 
Lines depict $g_l/P$ averaged over 500 simulation runs.
}
\label{fig:4}
\end{figure*}

The good agreement of the rotationally averaged order parameter profile $\langle \phi \rangle_{\varphi}$ and the
axisymmetric $\phi_{\mathrm{axi}}$ from our earlier work, both plotted in fig.\ \ref{fig:2} (b), suggests that in the steady swimming
state,  the full three-dimensional solution is also nearly axisymmetric about the swimming axis  $\mathbf{e}$.
However, in non-steady state we expect $\phi$ to deviate from axisymmetry, which we quantify by introducing 
an appropriate measure for the biaxiality of the order parameter field $\phi$.
In analogy to characterizing the orientational order of liquid crystals, we define
for the order parameter profile the traceless quadrupolar tensor \cite{mottram2014}
\begin{equation}
 \mathbf{Q}=\iint \phi\, \left(\mathbf{n}\otimes\mathbf{n}-\frac 1 3 \mathbbm{1}\right)\, \md\Omega\;,
\end{equation}
with surface normal $\mathbf{n}$, unit tensor $\mathbbm{1}$, and the
surface integral is performed over the whole droplet interface. 
Just as in the case of the moment of
inertia tensor, the eigenvalues and eigenvectors of $\mathbf{Q}$ characterize the symmetries of the order parameter field $\phi$. 
If two eigenvalues of $\mathbf{Q}$ are equal, $\phi$ is said to be uniaxial. On the other hand, if all eigenvalues of $\mathbf{Q}$ are distinct, $\phi$ is biaxial. Finally, the case of three vanishing eigenvalues, \emph{i.e.}, $\mathbf{Q}=0$, describes an isotropic or uniform order parameter field $\phi$ or at least with tetrahedral or cubic symmetry.
A measure for the degree of biaxiality, which incorporates the three mentioned cases, is given by
the biaxiality parameter \cite{longa1990,kaiser1992}
\begin{equation}
 \Delta=1-6\frac{(\mathrm{tr}\mathbf{Q}^3)^2}{(\mathrm{tr}\mathbf{Q}^2)^3}\;.
\label{eq:biaxiality}
\end{equation}
If the order parameter field $\phi$ is axisymmetric or isotropic, $\Delta=0$, while with increasing 
biaxiality $\Delta$ approaches $1$. 

In fig.\ \ref{fig:2} (a), we plot $\Delta$ as a function of time. At the initial time $t=0$, the order parameter profile is roughly 
uniform with $\Delta \approx 0$ (not visible). As the droplet speeds up, the 
biaxiality parameter $\Delta$ fluctuates strongly between $0$ and $1$. Starting at $t\approx 3$, $\Delta$ 
sharply decreases towards zero before the swimming speed becomes maximal.
Finally, in the steady swimming state, $\Delta$ 
is nearly zero but still fluctuates due to the thermal noise in the
order parameter profile $\phi$, which we indicate by
the error bars in fig.\ \ref{fig:2} (b). Hence, during the speed up of the droplet, the order parameter field $\phi$ clearly
is not axisymmetric.

\subsection{Coarsening dynamics}\label{subsec:coarsening_dynamics}

The period of strong biaxiality goes in hand with the coarsening dynamics of the order parameter profile towards
steady state.
Figure\ \ref{fig:4}(a) shows the order parameter profile $\phi(\theta,\varphi)$ 
at various time steps for the same simulation 
run as in fig.\ \ref{fig:2}.
Shortly after the simulation starts with the 
nearly uniform initial condition, small islands or domains
with $\phi>\phi_{\mathrm{eq}}$ and $\phi<\phi_{\mathrm{eq}}$ emerge,
which rapidly grow until $t \approx 1$, where the droplet hardly moves, see fig.\ \ref{fig:2} (a).
Then the coarsening or demixing process is slowed down. The domains coalesce on larger scales and the
droplet speeds up significantly.
Since the droplet interface area is finite, the domains
coalesce at some point to one large region which covers about half of the interface. From then on the droplet interface is covered by only two regions with $\phi<\phi_{\mathrm{eq}}$ and $\phi>\phi_{\mathrm{eq}}$.
\reB{The domain wall between the two regions is 
close to the equator and the}
droplet has reached its top speed [compare $v_D(t\approx 5)$ in fig.\ \ref{fig:2}(a)]. 
\reB{Then,}
the domain wall moves
\reB{towards the southern pole}
to its final position.
\reB{Since its circumference shrinks,}
the droplet speed $v_D$ slows down to 
its stationary value, which it reaches at $t\approx 9$. 
\reB{Thus, overshoot of the swimming velocity in fig.\ \ref{fig:2}(a) is the result of two processes taking place
on different time scales: the coarsening process and the final positioning of the domain wall at a somewhat larger time scale,
which depends on the parameters.}

Note that depending on the final position of the 
domain wall separating
the two regions, the droplet is either a pusher or a puller. 
If the domain wall with
increasing $\phi$ is situated in the southern hemisphere ($\pi/2 < \theta < \pi$), the droplet is a 
pusher. If it is located in the northern hemisphere ($0 < \theta < \pi/2$), a puller is realized.
\reB{However,} in our simulations the \reB{swimming} droplet is always a pusher
\reB{irrespective of the system parameters. This is due to the fact that the advective Marangoni current $\mathbf{j}_A$ at the interface
of the swimming droplet is always directed
towards the southern hemisphere. This flow also moves the domain wall away from the equator and towards the posterior end 
of the droplet.}
The squirmer parameter $\beta$ varies in 
the range $-2 < \beta < 0$ depending on Marangoni number $M$ (see fig.\ \ref{fig:3})
and equilibrium order parameter $\phi_{\mathrm{eq}}$. 
This is in agreement with earlier observations in ref.\ \cite{schmitt2013}. The timeframe $t>9$, where 
the swimming speed fluctuates around its steady-state value,
will be covered in sect.\ \ref{sec:dynamics_of_the_swimming_state}.

To quantify further the spatial structure of the order parameter profile during coarsening, we examine the angular
power spectrum $|s_l^m|^2$ of the surface tension.
It is related to $\phi$ in eq.\ (\ref{eq:sigma}).
Using the orthonormality relation
of spherical harmonics $Y_l^m(\theta,\varphi)$, given in appendix \ref{sec:spherical_harmonics}, one can 
compute the total power $P$ of the surface tension $\sigma$:
\begin{equation*}
 P=\iint  \sigma ^2\md\Omega = \sum_{l=1}^{\infty} g_l  =  \sum_{l=1}^{\infty} \limits \sum_{m=-l}^{l} |s_l^m|^2 \; .
 \end{equation*}
Here, the polar power spectrum $g_l$ characterizes the variation of the surface tension and thus the order parameter field $\phi$
along the polar angle $\theta$.
In particular, $g_l$ for small $l$ quantifies the large-angle variations of $\sigma$.
Note that $g_1$ is directly related to the swimming speed $v_D$ 
calculated from eq.\ (\ref{eq:velocity_vector}) in the polar coefficients $s_1^m$. 
Using $s_1^{-1} = -\overline{s}_1^1$, we find $g_1 = 3 \pi [(2+3\nu) v_D]^2$.

Figure\ \ref{fig:4}(b) depicts the polar power spectrum $g_l$ normalized by
the total power $P$ at the same time steps of the coarsening dynamics
discussed before in Fig.\ \ref{fig:4}(a). 
We also show an ensemble average of $g_l/P$.
At the initial time $t=0$, the spectrum of $g_l$ is solely characterized by 
frequencies or polar contributions
of the noisy initial condition $\phi(t=0)=\phi_{\mathrm{eq}} + \delta\phi$. Thus, the maximum 
frequency or polar number $l$ of the spectrum at $t=0$ is set by the level of refinement of the simulation mesh. 
During the initial period of fast coarsening until $t=1$,
the polar power spectrum shifts
from high to low frequencies indicating the increase of domain sizes.
Then the higher frequencies vanish more and more from the spectrum, 
as the phases associated with $\phi<\phi_{\mathrm{eq}}$ and $\phi>\phi_{\mathrm{eq}}$ separate.
Eventually, the spectrum $g_l$ strongly peaks at $l=1$ while the remaining coefficients 
become insignificant in comparison. Finally, from $t=5$ to $t=9$,  the first coefficient $g_1$ of the angular power spectrum 
decreases again while the second and third coefficients $g_2$ and $g_3$ rise. 
This confirms that in the final stage the droplet slows down its velocity $v_D$ and 
tunes its squirmer parameter $\beta$ by shifting the domain wall further away from the equator.

\begin{figure}
\includegraphics[width=0.34\textwidth, angle=-90,trim=0mm 0mm 0mm 0mm, clip]{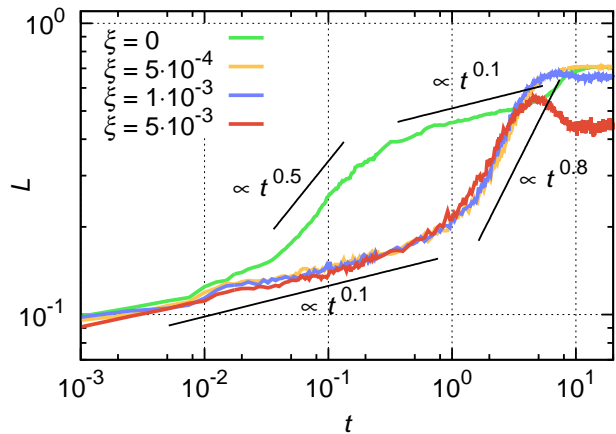}
\caption{
Mean domain size $L$ averaged over 200 simulation runs plotted versus
reduced time in units of $\tau_D$ for different noise strengths $\xi$.
A domain is defined by a compact region with $\phi>\phi_{\mathrm{eq}}$. 
Same parameters as in fig.\ \ref{fig:4} are used.
}
\label{fig:5}
\end{figure}

In order to quantify further the temporal evolution
of the coarsening dynamics, we will now 
investigate the average domain size as a function of time.
We define the mean linear size of a phase domain by
\begin{equation*}
 L=\sqrt{\frac{\langle v_n^+\rangle}{v_n}}\; .
\end{equation*}
Here, $\langle v_n^+\rangle$ denotes the averaged number of grid points in a connected region, where $\phi$ is larger than 
$\phi_{\mathrm{eq}}$, and $v_n$ is the total number of grid points.
Thus, the domain size lies within the range $\sqrt{1/v_n}\leq L \leq 1$, and 
$L(t)$ should increase during the coarsening dynamics towards the steady swimming state. 
The fluctuations $\delta\phi$ of the initial profile
are normal distributed with zero mean such that at $t=0$ half of the grid points 
have $\phi > \phi_{\mathrm{eq}}$.
They cannot all be isolated but rather belong to small connected regions with $L\approx \sqrt{5/v_n}$,
where we extracted the factor $\sqrt{5}$ from our simulations at $t=0$.
Furthermore, we expect the maximum length to be around $L\approx \sqrt{1/2}$. 
So, in our simulations $L(t)$ lies in the interval $\sqrt{5/v_n}\leq L \leq \sqrt{1/2}$.
Figure\ \ref{fig:5} shows $L(t)$ averaged over 200 simulation runs
for different noise strengths $\xi$. The other parameters are
the same as before.
We clearly see a separation of time scales of the coarsening dynamics
for both cases, with and without noise.
At early times, we find in both cases a power law behavior $L(t) \propto t^{0.1}$. 
Without noise, coarsening quickly speeds up at a rate $L(t) \propto t^{1/2}$ and then slows down again to 
$L(t) \propto t^{0.1}$.
In contrast, thermal fluctuations in the order parameter profile hinder early coarsening and the mean domain size continues to grow
slowly with $L(t) \propto t^{0.1}$ over several decades and then crosses over to a fast final coarsening with rate $L(t) \propto t^{0.8}$.
The crossover time is only determined by the diffusion time $\tau_D$ and does not depend on noise strength $\xi$.
Interestingly, a similar observation to the second case
has been made for coarsening in the dynamical model H,
where the Cahn-Hilliard equation couples to fluid flow at low-Reynolds number via an advection term.
A slow coarsening rate $ L(t) \propto t^{1/3}$ in a diffusive regime at short times
is followed by an advection driven 
regime with $L(t) \propto t$ at later times
\cite{bray2003,bray2002,brenier2011,otto2013}. 
Although we cannot simply reformulate our model as an advective Cahn-Hilliard equation, since the phase separation
in our case is driven by the interfacial flow $\mathbf{u}|_R$ itself, we observe similar coarsening regimes as
in model H, when we include some noise.

\section{Dynamics of the swimming state}\label{sec:dynamics_of_the_swimming_state}

\begin{figure}
\includegraphics[width=0.34\textwidth, angle=-90, trim=0mm 0mm 0mm 0mm, clip]{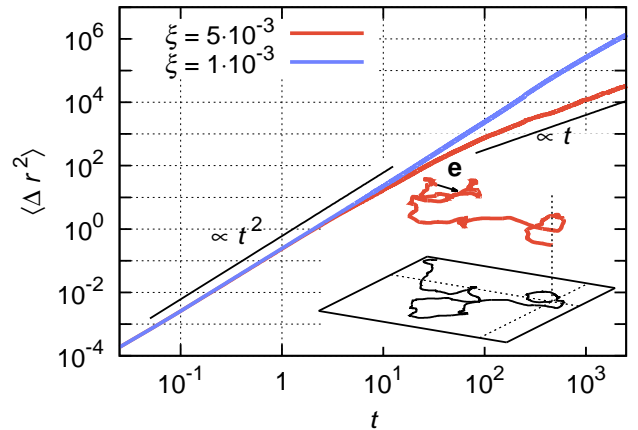}
\caption{Mean square displacement of the swimming active droplet
for different noise strengths $\xi$. At $t=0$ the droplet is already in the swimming state.
Inset: A typical trajectory $\mathbf{r}(t)$ of an active droplet subject to noise with 
strength $\xi=5\cdot 10^{-3}$. 
The trajectory is reminiscent of an active particle with constant speed and rotationally diffusing orientation vector $\mathbf{e}(t)$.
}\label{fig:6}
\end{figure}

We now consider the time regime $t >9$, where the droplet moves in its steady swimming state.
However, as can be observed in fig.\ \ref{fig:2} (a), the droplet speed 
$v_D(t>9)$ in the swimming state 
strongly fluctuates since we have added a thermal noise term to the
diffusion-advection-reaction equation\ (\ref{eq:conti_dimless})
for the order parameter field $\phi$.
These fluctuations also randomly change the swimming direction $\mathbf{e}$ as the inset of fig.\ \ref{fig:6} illustrates,
where we show an exemplary swimming trajectory $\mathbf{r}(t)=\mathbf{r}(0)+\int_0^t \md t' v_D(t')\mathbf{e}(t')$.
Therefore, we expect the droplet to perform active Brownian motion or a persistent random walk. In a droplet with 
axisymmetric profile the swimming direction is perpendicular to the domain wall separating both phases. When the
order-parameter profile fluctuates, we also expect the domain wall to fluctuate and thereby the swimming direction 
$\mathbf{e}$. There are no other reasons to change the orientation of $\mathbf{e}$. In ref.\ \cite{schmitt2016}
we showed that a spherical \reB{and isotropic} emulsion droplet, with Marangoni flow at its surface,
does not experience a frictional torque, which could also change the swimming direction. 
\reB{Thus, for an arbitrary surface tension profile $\sigma(\theta,\varphi)$ a spinning motion of the droplet does not occur.}
But this also means that fluctuating flow fields in the surrounding fluid, 
\reB{which have to fulfill boundary condition\ (\ref{eq.boundary}) at the droplet surface,}
cannot generate a stochastic torque acting on the droplet.
\reB{Therefore, in contrast to a rigid colloid, spherical emulsion droplets 
do not exhibit conventional thermal rotational diffusion.}

\subsection{Active Brownian motion of the droplet}\label{subsec:active_brownian_motion}

To characterize the active Brownian motion of the droplet, we first discuss the mean squared displacement (MSD) $\langle \mathrm{\Delta} r^2\rangle=\langle [\mathbf{r}(t)-\mathbf{r}(0) ]^2\rangle$, where we average over an ensemble of trajectories.
Here, the droplet is already in the swimming state at $t=0$, thus the MSD does not include
the droplet's acceleration towards the steady swimming state as discussed in sect.\ \ref{sec:dynamics_towards_the_swimming_state}.
Figure\ \ref{fig:6} shows the MSD for a droplet with noise strength $\xi=5\cdot 10^{-3}$.
At early times, the droplet moves ballistically since the MSD 
grows as $\langle \mathrm{\Delta} r^2\rangle \propto t^2$, while between $t= 10$ and $t=100$ 
it crosses over to diffusive motion with $\langle \mathrm{\Delta} r^2\rangle \propto t$. This motion persists as $t\rightarrow\infty$.
As expected, in the absence of noise, $\xi=0$, we 
always observe ballistic motion $\langle \mathrm{\Delta} r^2\rangle \propto t^2$ (not shown).
The MSD for $\xi = 10^{-3}$ in fig.\ \ref{fig:6} does not cross over to diffusion in the plotted time range.
In the following, we will discuss the
influence of the noise strength $\xi$ on the Brownian motion in more detail but we will first 
introduce 
what has become the standard model of an active Brownian particle \cite{howse2007,lobaskin2008,enculescu2011,romanczuk2012}.

If we assume the droplet speed $v_D$ and orientation vector $\mathbf{e}$ to be independent random variables,
we can factorize the MSD as 
\begin{equation*}
\langle \mathrm{\Delta} r^2\rangle =\int_0^t\limits\md t'\int_0^t\limits\md t'' \langle v_D(t') v_D(t'')\rangle \langle\mathbf{e}(t')\cdot\mathbf{e}(t'')\rangle\;.
\end{equation*}
For active Brownian particles without any aligning field the swimming direction diffuses freely on the unit sphere,
which one describes by the rotational diffusion equation $\partial_t p(\mathbf{e},t)=D_r\laplaces p(\mathbf{e},t)$.
Thus the orientational correlation function decays as \cite{lovely1975,doi1988}
\begin{equation}
\langle \mathbf{e}(0)\cdot\mathbf{e}(t)\rangle=\me^{-t/\tau_r}\;.\label{eq:rotational_corr_fun}
\end{equation}
Here, the rotational correlation time $\tau_r=1/(2D_r)$ is the characteristic time it takes the droplet to ``forget" about 
the initial orientation $\mathbf{e}(0)$. 
Hence, for times $t<\tau_r$ the droplet swims roughly in the direction of $\mathbf{e}(0)$, while at later times 
$t>\tau_r$ the orientation becomes randomized.

Under the assumption of a constant swimming speed, \emph{i.e.} $\langle v_D(t') v_D(t'')\rangle=(v_D)^2$, one finds for the MSD
\begin{equation}
 \langle \mathrm{\Delta} r^2\rangle=2(v_D\tau_r)^2\left(\frac t {\tau_r}-1+\me^{-t/\tau_r}\right)\;.\label{eq:msd}
\end{equation}
Expression (\ref{eq:msd}) confirms the findings of fig.\ \ref{fig:6}: Ballistic motion $\langle \mathrm{\Delta} r^2\rangle=(v_D t)^2$ 
with velocity $v_D$ at $t\ll \tau_r$ and diffusive motion with 
\begin{equation}
\langle \mathrm{\Delta} r^2\rangle = 6 D_{\mathrm{eff}} t
\quad
\mathrm{and} 
\quad
D_{\mathrm{eff}}=(v_D)^2\tau_r/3 \label{eq:D_eff_tau}
\end{equation}
for $t\gg \tau_r$. Here, $D_{\mathrm{eff}}$ is the effective translational diffusion constant.
It neglects any contribution from thermal translational motion, which is o.k. 
for sufficiently large $v_D$.

\begin{figure}
\includegraphics[width=0.33\textwidth, angle=-90, trim=0mm 0mm 0mm 0mm, clip]{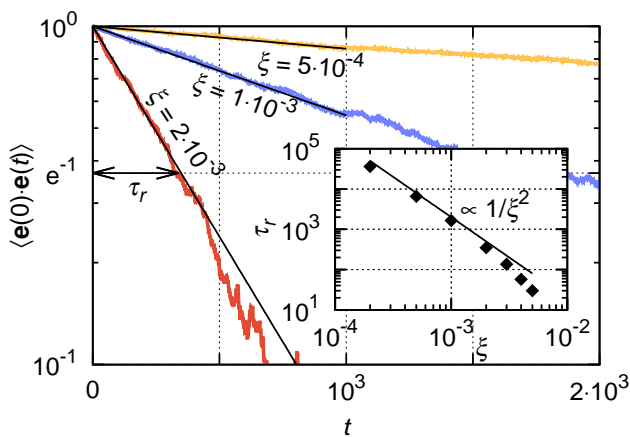}
\caption{Rotational correlation function of the active droplet and fits to
$\me^{-t/\tau_r}$ for different values of noise strength $\xi$.
At the rotational correlation time $\tau_r$, one has
$\langle \mathbf{e}(0)\cdot\mathbf{e}(\tau_r)\rangle=\me^{-1}$, as illustrated for the case $\xi=2\cdot 10^{-3}$.
Inset: $\tau_r$ plotted versus noise strength $\xi$ and a fit to $\xi^{-2}$.
}
\label{fig:7}
\end{figure}

Indeed, for the active droplet the rotational correlation function $\langle \mathbf{e}(0)\cdot\mathbf{e}(t)\rangle$
decays exponentially as demonstrated in fig.\ \ref{fig:7}
for different noise strengths and by fits to eq.\ (\ref{eq:rotational_corr_fun}).
The rotational correlation time $\tau_r$, which acts as 
fitting parameter, is shown in the inset for various values of noise strength $\xi$. 
For $\xi=5\cdot 10^{-3}$, we find $\tau_r\approx 30$, which is in agreement with the 
cross-over region from ballistic to diffusive motion in the MSD curve of fig.\ \ref{fig:6}. 
Furthermore, from the asymptotic behavior at
$t\ll \tau_r$ and $t\gg\tau_r$ of the MSD in fig.\ \ref{fig:6}, we find $v_D\approx 0.3$ and $D_{\mathrm{eff}}\approx 1$, respectively. This gives the rotational correlation time $\tau_r= 3D_{\mathrm{eff}}/(v_D)^2\approx 33$, which is close to the value determined from the orientational correlations.
Thus $D_{\mathrm{eff}}$ and $\tau_r$ comprise the same information about the droplet trajectory $\mathbf{r}(t)$. 
However, the measurement of $\tau_r$ in experiments or simulations can be done on much shorter time scales than $D_{\mathrm{eff}}$.
\reB{Note that the relative fluctuations of the swimming speed about its mean value are small, as fig.\ \ref{fig:2}(a) demonstrates. 
Therefore, they do not have a strong effect on $D_{\mathrm{eff}}$ and we can safely use the mean value $v_D$ in 
eq.\ (\ref{eq:D_eff_tau}).}

We do not know published experimental data for trajectories of active droplets in an unbounded fluid.
However, fig.\ 1 of ref.\ \cite{thutupalli2011} shows a trajectory of an active droplet confined between two glass plates. 
One can estimate the rotational correlation time $\tau_r$ to be on the order of $100\mathrm{s}$.
To compare this value with our model, we recapitulate the noise strength $\xi=\ell / R$, which connects surfactant head size $\ell$ 
with droplet radius $R$, see sect.\ \ref{subsec:system_parameters}. 
If we assume, $\xi\approx 10^{-4}\dots 10^{-3}$, we find from fig.\ \ref{fig:7} a rotational correlation time $\tau_r\approx 10^4$
given in units of diffusion time $\tau_D=R^2/D$ with interfacial diffusion constant $D$. Typical values for $D$ are on the order of
$10^{-5} \mathrm{cm}^2/\mathrm{s}$ \cite{dukhin1995}. Thus, for a droplet with $R$ on the order of $10\mathrm{\mu m}$, one 
finds $\tau_D\approx 0.1 \mathrm{s}$
and the rotational correlation time 
$\tau_r\approx 10^3\mathrm{s}$.

This is only a factor 10 larger than the
estimated value of $100\mathrm{s}$ from ref. \cite{thutupalli2011}.
Given some uncertainties in our estimate such a difference can be expected. Nevertheless, two causes for the discrepancy are thinkable. First and foremost, our model droplet is allowed to move freely in the bulk fluid, 
while the real droplet of ref. \cite{thutupalli2011} is confined between two plates, which limits the degrees of freedom and thus alters $\tau_r$.
Secondly, active emulsion droplets are usually immersed in a surfactant laden fluid well above the critical micelle concentration. Hence, the 
surfactants from the bulk adsorb in form of micelles. This leads to local disturbances in the surfactant mixture at the front of the swimming droplet, and hence to an additional randomization of the droplet trajectory. We recently modeled the adsorption of micelles 
explicitly in a different system\ \cite{schmitt2016}.

\subsection{How fluctuations randomize the droplet direction}\label{subsec:role_of_fluctuations}

Now, we develop a theory how the noise strength $\xi$ influences the rotational diffusion of the droplet direction.
By increasing $\xi$ in the diffusion-advection-reaction equation\ (\ref{eq:conti_dimless}), the order parameter profile $\phi$ is subject 
to stronger fluctuations. 
In particular, these fluctuations 
affect shape and orientation of the 
domain wall separating the two regions with
$\phi<\phi_{\mathrm{eq}}$ and $\phi>\phi_{\mathrm{eq}}$
from each other.
The surface flow field is largest in this domain wall and thereby the orientation of the wall on the droplet interface 
determines the droplet swimming vector $\mathbf{e}$.
Thus, increasing noise strength $\xi$ results in 
stronger fluctuations of $\mathbf{e}$ and ultimately a
more pronounced rotational diffusion.  
The inset of fig.\ \ref{fig:7} confirms this 
scenario for the rotational correlation time $\tau_r$. 
Interestingly, for noise strengths up to $\xi \approx 3\cdot 10^{-3}$, 
one fits the data quite well by $\tau_r\propto 1/\xi^{2}$.
\reA{Since the noise strength $\xi$ was defined as $\xi=\ell/R$ in sect.\ \ref{subsec:system_parameters}, 
the rotational diffusion constant $D_r=1/(2\tau_r)$ of the active droplet behaves as $D_r\propto 1/R^2$, which is in contrast to the scaling $D_r\propto 1/R^3$ of a passive colloid. In addition, one finds for the total number of surfactants $N=4\pi R^2/\ell^2$  that $D_r\propto 1/N$. This can be understood from a simple hand-waving argument.}

\reA{Fluctuations in the order parameter profile described in eq.\ (\ref{eq:conti_dimless}) correspond to  
exchanging surfactant molecules (brominated against non-brominated and vice versa).
A single event initiates an angular displacement $\Delta \varphi \approx 2\pi /\sqrt{N}$ 
of the droplet direction $\mathbf{e}$. These fluctuations take place on the diffusive time scale $\tau$. Furthermore, from eq.\ (\ref{eq:rotational_corr_fun}) one finds a diffusive mean squared angular displacement $\langle (\Delta \varphi)^2\rangle=4D_r t$ for times $t\ll\tau_r$. Thus, for $t$ on the time scale $\tau$, one finds $D_r \propto 1/N$.}
\reA{In what follows, we want to explain the scaling $\tau_r\propto 1/\xi^{2}$ more rigorously}
by
applying perturbation theory to the thermal fluctuations of the order parameter profile around its
steady profile.

\begin{figure}
\includegraphics[width=0.48\textwidth, angle=0, trim=0mm 0mm 0mm 0mm, clip]{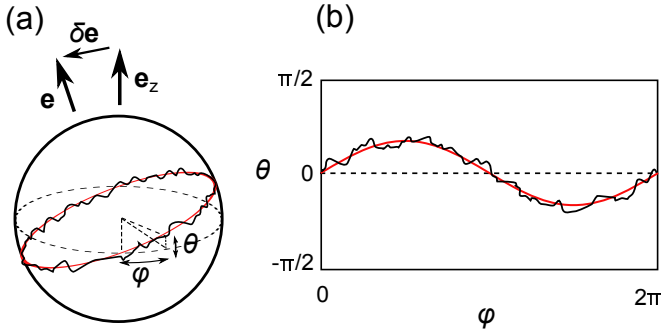}
\caption{Illustration of a reorienting droplet. (a) The black curve around the droplet interface shows the noisy phase boundary denoted in the coordinate system $(\theta,\varphi)$ of the droplet without noise. The red curve shows the first mode of a Fourier expansion, see text.
(b) Flat representation in the said coordinate system $(\theta,\varphi)$.}
\label{fig:8}
\end{figure}
As mentioned before, small fluctuations of the domain wall result in
random changes of the droplet direction. Figure\ \ref{fig:8} shows an exaggerated illustration of the situation. 
Plot (a) illustrates a tilt in the orientation of the domain wall generated by the sinusoidal variation of the polar angle $\theta$
along the azimuthal angle $\varphi$.
In general, fluctuations of the domain wall can be 
decomposed into Fourier modes,
$\theta=\sum_m a_m \sin[m(\varphi-\varphi_{0m})]$.
Only the first mode, $m=1$, of this expansion determines the change in orientation, $\delta \mathbf{e}$, as illustrated in fig.\ \ref{fig:8} (a). 
All higher modes cannot change the swimming direction since the effects of the resulting 
surface flow field on $\mathbf{e}$ cancel each other. 

We now apply perturbation theory to the fluctuating order parameter profile, which determines the surface tension profile
and thereby the swimming direction according to eq.\ (\ref{eq:velocity_vector}). We
consider a droplet, which initially swims in $z$-di\-rec\-tion and changes its direction in $x$ and/or $y$-di\-rec\-tion,
hence $\mathbf{e}=\mathbf{e}_z + \delta \mathbf{e}$.
We write down a perturbation ansatz for the surface tension profile, $\sigma= \sigma_0 + \delta \sigma$,
with the unperturbed axisymmetric part $\sigma_0=\sum_{l=1}^{\infty} s_l^0 Y_l^0$ and the perturbation 
$\delta \sigma= s_1^1 Y_1^1+s_1^{-1} Y_1^{-1}$, where we only include the coefficients $s_1^{\pm 1}$, which are 
responsible for changes $\delta \mathbf{e}$, as one recognizes from eq.\ (\ref{eq:velocity_vector}).
By linearizing the equation of state (\ref{eq:sigma}) around $\phi_{\mathrm{eq}}$, one can connect the coefficients $s_l^m$ of $\sigma$ directly to the expansion coefficients of the order parameter field
$\phi$. Writing $\phi=\phi_0+\delta\phi$, where $\phi_0$ describes the unperturbed steady-state field and $\delta \phi$ its fluctuations,
we find $\phi_0 = a \sigma_0$ and $\delta \phi = a \delta \sigma$, where the factor $a$ is given in 
appendix\ \ref{sec:perturbation_ansatz}. Similarly, one decomposes $\mathbf{j}_D$ and $\mathbf{u}|_R$ into their steady-state
fields and a fluctuating small perturbation (see appendix\ \ref{sec:perturbation_ansatz}).
This allows us to derive from the field equation (\ref{eq:conti_dimless}) of the order parameter, the dynamic equation linear in
the fluctuating perturbations:
\begin{equation}
   \partial_{t}\delta\phi=-\nablas\cdot\left[\delta\mathbf{j}_D + M\left(\delta\phi\mathbf{u}_{0}+\phi_0\delta\mathbf{u}\right)\right]
   - \kappa \delta\phi+\xi\zeta\;.
      \label{eq:first_order_field_equation}
\end{equation}
From our study of the coarsening dynamics we know that the first and second term on the right-hand side describe a relaxation
towards steady state on times $t <10$. The rotational diffusion of the droplet direction occurs on time scales much larger and can only
be due to the noise term. Extracting from Eq.\ (\ref{eq:first_order_field_equation}) the coefficients $s_1^{\pm 1}$ relevant for 
$\delta \mathbf{e}$, we obtain
\begin{equation}
   \partial_{t}s_1^{\pm 1} \simeq \frac{\xi}{a}\zeta_1^{\pm 1}\; . \label{eq:projected_perturbation}
\end{equation}
A more thorough derivation of Eq.\ (\ref{eq:projected_perturbation}) is presented in appendix\ \ref{sec:projection_onto_s_1}.
We have decomposed noise $\zeta$ into its multipole moments, $\zeta=\sum_{l,m}\zeta_l^m Y_l^m$. Projecting the variance of
eq.\ (\ref{eq:fluc_diss_b_dimless}) onto the relevant spherical harmonics, we obtain
the fluctuation-dissipation theorem
\begin{equation}
 \langle \zeta_l^m(t) \overline{\zeta}_{l'}^{m'}(t')\rangle=2l(l+1)\delta(t-t') 
 \delta_{l,l'} \delta_{m,m'}   \;.\label{eq:fluc_diss_components}
\end{equation}

Assuming a constant speed $v_D$ during the reorientation of the droplet, we use
eq.\ (\ref{eq:projected_perturbation}) 
in eq.\ (\ref{eq:dimless_velocity_vector}) for the droplet velocity vector to formulate the
stochastic equation for rotations of the direction vector $\mathbf{e}$:
\begin{equation}
 \partial_t\mathbf{e}=\frac{\xi}{\sqrt{6\pi}v_D (2+3\nu)a}\delta\boldsymbol{\zeta}\;,\label{eq:langevin_rot}
\end{equation}
where we introduced the rotational noise vector
\begin{eqnarray*}
\delta\boldsymbol{\zeta}=\left(\begin{array}{c} \zeta_1^1-\zeta_1^{-1} \\i\left(\zeta_1^1+\zeta_1^{-1}\right)  \\ 0 \end{array}\right)\;.
\end{eqnarray*}
By comparing eq.\ (\ref{eq:langevin_rot}) with the Langevin equation 
for the Brownian motion of a particle's orientation $\mathbf{e}$ due to 
rotational noise $\boldsymbol{\eta}_r$:  
$\partial_t \mathbf{e}=\sqrt{2D_r} \boldsymbol{\eta}_r \times \mathbf{e}$ \cite{dhont1996},
we identify $\delta\boldsymbol{\zeta}=\boldsymbol{\eta}_r\times \mathbf{e}$ and
\begin{equation}
\frac{\xi}{\sqrt{6\pi}v_D(2+3\nu)a}=\sqrt{2D_r}\;.
\end{equation}
Hence, the rotational correlation time $\tau_r=1/(2D_r)$ scales as $\tau_r\propto 1/{\xi^2}$ with noise strength $\xi$. This confirms the fit in the inset of fig. \ref{fig:7}
\reB{for noise strengths up to $\xi\approx 10^{-3}$. For larger
$\xi$, the fluctuations start to very strongly disturb the domain wall. 
The illustration of fig.\ \ref{fig:8} is no longer valid and with it the the perturbation theory breaks down.
Instead, the droplet loses its persistent swimming axis and the motion becomes purely erratic, which manifests itself in a rapidly decreasing 
$\tau_r$.}

Thus, beyond the time scale, the order parameter profile needs to reach its steady state, \reB{and for $\xi<10^{-3}$,}
the dynamics of the swimming active emulsion droplet 
is equivalent to the dynamics of an active Brownian particle with constant swimming velocity and rotationally diffusing orientation vector $\mathbf{e}$. 

\section{Conclusions}\label{sec:conclusions}

In this paper we considered an active emulsion droplet, which is driven by solutocapillary Marangoni flow at its 
interface \cite{thutupalli2011}. A diffusion-advection-reaction equation for the surfactant mixture at the droplet interface, which 
we formulated in ref. \cite{schmitt2013}, is used together with the analytic
solution of the Stokes equation \cite{schmitt2016}. 
By omitting the axi\-symmetric constraint and including thermal noise into the description of the surfactant mixture,
we generalized the model of ref. \cite{schmitt2013} to a full three-dimensional system and  
thereby were able to focus on new aspects.

First, we explored the dynamics from a uniform, but slightly perturbed surfactant mixture to the uniaxial steady swimming state,
where the two surfactant types are phase-separated.
In between the initial and the swimming state, the surfactant mixture is not axisymmetric, which we verified by introducing and 
evaluating a biaxiality measure. We then investigated in detail the coarsening dynamics 
towards the swimming state by means of the polar power spectrum of the surface tension $\sigma$ as well as the average domain 
size of the surfactant mixture.
The coarsening proceeds in two steps. An initially slow growth of domain size is followed by a nearly ballistic regime, which 
is reminiscent to coarsening in the dynamic model H \cite{bray2003}.

Second, we studied the dynamics of the squirming droplet. Due to the included thermal noise,
the surfactant composition fluctuates and thereby the droplet constantly changes its swimming direction 
performing a persistent random walk.
Thus, the swimming dynamics of the squirming droplet is a typical example of an active Brownian particle.
The persistence of the droplet trajectory depends on the noise strength $\xi$.
It is characterized by the rotational correlation time,
for which we find the scaling law $\tau_r\propto\xi^{-2}$.
In fact, we are able to explain this scaling by applying perturbation theory to the diffusion-advection-reaction equation for 
the mixture order parameter.
Thus we can link the dynamics of the surfactants at the molecular level to the dynamics of the droplet as a whole.

\reB{We hope that our work initiates further research in the field of
active emulsion droplets. A deeper theoretical understanding of the coarsening due to the 
Marangoni effect could help to understand the power laws that we found in our simulations.
Furthermore, various extensions of this work are possible,
\emph{e.g.}, the explicit implementation of micellar adsorption as discussed in ref.\ \cite{schmitt2016} or 
taking into account
confining plates below and above the droplet via no-slip boundary conditions. 
Finally, a numerical study of the collective motion of active droplets, which swarm in experiments\ \cite{thutupalli2011}, is still missing in the literature 
but has been implemented for pure squirmers \cite{zoettl2014}.
}

Exploring and understanding the swimming mechanisms of both biological and artificial microswimmers is one of the challenges
in the field. Here, we demonstrated that this task involves new and fascinating physics. Having gained deeper insights into
these mechanisms can help to further improve the design of artificial microswimmers and tailor them for specific needs such 
as cargo transport.

\begin{acknowledgement}
We acknowledge financial support by the Deutsche Forschungsgemeinschaft in the framework of the collaborative research 
center SFB 910, project B4 and the research training group GRK 1558.
\end{acknowledgement}

\begin{appendix}

\section{Spherical harmonics}\label{sec:spherical_harmonics}

Throughout this paper we use the following definition of spherical harmonics:
\begin{equation*}
Y_l^m( \theta , \varphi ) = \sqrt{\frac {2l+1}{4\pi}\frac{(l-m)!}{(l+m)!}} \, P_l^m ( \cos{\theta} ) \, \me^{i m \varphi}\;,
\end{equation*}
with associated Legendre polynomials $P_l^m$ of degree $l$, order $m$, and with orthonormality:
\begin{equation*}
 \iint Y_l^m \, \overline{Y}_{l'}^{m'} \md\Omega =\delta_{l,l'}\, \delta_{m,m'}\;,
\end{equation*}
where $\overline{Y}_l^m$ denotes the complex conjugate of $Y_l^m$.

The spherical harmonics fulfill the following helpful relations:
\begin{subequations}
\label{eq:integrals}
\begin{eqnarray}
 \iint Y_l^0 Y_1^m \overline{Y}_{1}^{m'} \md\Omega	&=&\frac{-1}{\sqrt{20\pi}}\delta_{l,2} \delta_{m,m'}\;, \\
 \iint \nablas Y_l^0\cdot\nablas Y_1^m \overline{Y}_{1}^{m'} \md\Omega&=& \frac{-3}{\sqrt{20\pi}}\delta_{l,2} \delta_{m,m'}\;,
\end{eqnarray}
\end{subequations}
where $\nablas$ is the directional gradient defined in sect.\ \ref{subsec:diffusion-advection-reaction_equation} 
and evaluated at $r=1$.

\section{Squirmer parameter}\label{sec:beta_details}

The squirmer parameter for a droplet swimming in an arbitrary direction is given by \cite{schmitt2016}:
\begin{subequations}\label{eq:beta_full}
\begin{eqnarray}
 \beta&=&-\sqrt{\frac {27} 5} \frac {\tilde s_2^0}{|\tilde s_1^0|} \,,\label{eq:beta_full_1}\\
 \tilde s_1^0 &=& \sqrt{(s_1^0)^2-2s_1^1s_1^{-1}} \;,\label{eq:beta_full_2}\\
 \tilde s_2^0 &=& \bigg(\!\sqrt{6} \left[s_2^2(s_1^{-1})^2+s_2^{-2}(s_1^1)^2\right]\!-\!\sqrt{12}s_1^0\left[s_2^1s_1^{-1}\!+\!s_2^{-1}s_1^1\right]\notag\\
&&+2s_2^0\left[(s_1^0)^2+s_1^1s_1^{-1}\right]\bigg)\bigg / \left[2(s_1^0)^2-4s_1^1s_1^{-1}\right]\label{eq:beta_full_3}\;,
\end{eqnarray}
\end{subequations}
with coefficients $s_l^m$ from eq.\ (\ref{eq:s_l^m}). By setting $m=0$, this reduces to the case of an axisymmetric droplet swimming along the $z$-direction.

\section{Element noise vector}\label{sec:element_moise_vector}

Here, we discretize the thermal
noise $\zeta$ in eq.\ (\ref{eq:conti_dimless}) 
and obtain the element noise vector $\underline{z}$
with component $z_i$ for the FVM element $i$. 
We define the correlation function between $z_i$ and $z_j$ by integrating eq.\ (\ref{eq:fluc_diss_b_dimless}) over element areas $A_i$ and $A_j$:
\begin{subequations}
\begin{eqnarray}
&&\langle z_i(t)z_j(t')\rangle\equiv\iint\limits_{A_i}\! \md A_i \iint\limits_{ A_j}\! \md A_j \langle\zeta(\mathbf{r}_i,t)\zeta(\mathbf{r}_j,t')\rangle\quad\\
&&=2\int_{\partial A_i}\limits\! \md S_i\, \mathbf{n}_i\cdot\!\int_{\partial A_j}\limits\! \md S_j\mathbf{n}_j\,\delta(\mathbf{r}_i-\mathbf{r}_j)\delta(t-t')\label{eq:element_noise_vector2}\\
&&=2 \sum_{q} l_{iq}\sum_{p} l_{jp}\delta_{q,p}\mathbf{n}_{iq} \cdot \mathbf{n}_{jp} \delta(t-t')\;.\label{eq:element_noise_vector3}
\end{eqnarray}
\end{subequations}
In eq.\ (\ref{eq:element_noise_vector2}) we used the divergence theorem and in eq.\ (\ref{eq:element_noise_vector3}) we converted 
the line integrals into sums over the element boundaries.
Furthermore, we discretized $\delta(\mathbf{r}_i-\mathbf{r}_j)$ by partitioning the surface into rhombi of area $A_{\Diamond}$ (see fig.\ \ref{fig:1}) and 
defined
\begin{equation*}
 \delta_{q,p}= \begin{cases} 
1/A_{\Diamond} & \mathrm{for} \enspace q=p\;, \\ 
0 & \mathrm{for} \enspace q\neq p\;,
\end{cases}
\end{equation*}
where $q$ and $p$ are the indices of the respective
boundaries of elements $i$ and $j$.
Three cases have to be considered.
First, if the elements $i$ and $j$ are neither identical nor neighbors, $\delta_{q,p}$ vanishes in eq.\ (\ref{eq:element_noise_vector3}) for 
all $q$ and $p$.
Second, for $i=j$,
$\delta_{q,p}=1/A_{\Diamond}$ and $\mathbf{n}_{iq} \cdot \mathbf{n}_{jp}=1$ for all $q$ and $p$. Finally, for neighboring elements 
there is one common boundary, where $\delta_{q,p}=1/A_{\Diamond}$ and $\mathbf{n}_{iq} \cdot \mathbf{n}_{jp}=-1$. Thus, one finds:
\begin{eqnarray}
\langle\underline{z}(t)\otimes\underline{z}(t')\rangle&=&\frac {2Nl^2}{A_{\Diamond}} \left(\underline{\underline{1}}-\frac 1 N \underline{\underline{Q}}\right) \delta(t-t')\;,
\label{eq:fluc_diss_element_conti}
\end{eqnarray}
where $N$ is the number of element boundaries. Here, $Q_{ij}=1$ if elements $i$ and $j$ are neighbors and zero otherwise. Note that in eq.\ (\ref{eq:fluc_diss_element_conti}), we assumed 
the same edge
length $l$ and number of boundaries $N$
for all elements.
This is reasonable for a refined icosahedron with 642 FVM elements, as discussed in sect.\ \ref{sec:finite_volume_method}.
The form of eq.\ (\ref{eq:fluc_diss_element_conti}) acknowledges the conservation law for the noise\ \cite{hawick2010}. However, in simulations we did not observe any effect of the next--neighbor correlations and therefore simplified the noise to the expression (\ref{eq:fluc_diss_element}) in the main text. Furthermore, we 
take
$N=6$ and $A_{\Diamond}=\sqrt{3/4}l^2$, since our grid is mostly hexagonal, which explains the prefactor $\sqrt{2Nl^2/A_{\Diamond}}=2 \cdot 12^{1/4}$ in eq.\ (\ref{eq:discretize}),
when we redefine the noise vector by the following replacement,
$\underline{z} \rightarrow 2 \cdot 12^{1/4} \underline{z}$.

\section{Average over droplet interface}\label{sec:averages}

The average
\begin{equation*}
 \langle f \rangle_{\varphi} = \frac 1 {2\pi}\int f(\theta,\varphi)\ \md\varphi\;,
\end{equation*}
is taken over the azimuthal angle $\varphi$ in the coordinate frame whose $z$-axis is directed along the swimming direction $\mathbf{e}$. Here, the front of the moving droplet is at $\theta=0$. 

\section{Perturbation ansatz}\label{sec:perturbation_ansatz}

The zero and first-order contributions of $\phi=\phi_0+\delta\phi$, $\mathbf{j}_D=\mathbf{j}_{D,0}+\delta\mathbf{j}_D$, and $\mathbf{u}|_R=\mathbf{u}_0+\delta\mathbf{u}$ are given by:
\begin{subequations}\label{eq:perturbation_ansatz}
\begin{eqnarray}
\phi_0&=&a\sum_{l=1}^{\infty} s_l^0 Y_l^0\;,\\
\delta\phi&=&a\left(s_1^1 Y_1^1+s_1^{-1} Y_1^{-1}\right)\;,
\label{eq:perturbation_ansatz_b}\\
\mathbf{j}_{D,0}&=&-b\nablas\phi_0\;,\\
\delta\mathbf{j}_D&=&-b\nablas\delta\phi\;,\\
\mathbf{u}_0&=&c s_1^0 \nablas Y_1^0+\sum_{l=2}^{\infty}\limits\frac{s_l^0}{2l+1} \nablas Y_l^0\;,\\
\delta\mathbf{u}&=&c\left(s_1^1 \nablas Y_1^1+s_1^{-1}\nablas Y_1^{-1}\right)\;,
\end{eqnarray}
\end{subequations}
with parameters
\begin{subequations}\label{eq:constants}
\begin{eqnarray}
a&=&\frac{4(b_1-b_2)}{2(b_1-b_2)+\phi_{\mathrm{eq}}(b_1+b_2)}\approx 1.14\;,\\
b&=&(1-\phi_{\mathrm{eq}}^2)^{-1}-\frac 1 2 (b_1+b_2-b_{12})\approx 1.11\;,\\
c&=&({1+\nu})/({2+3\nu})\approx 0.49\;.
\end{eqnarray}
\end{subequations}
Here we used the values of sect.\ \ref{subsec:system_parameters} for $b_1, b_2, b_{12}, \phi_{\mathrm{eq}}$ and $\nu$. 

\section{Dynamic equation for $s_1^{\pm 1}$}\label{sec:projection_onto_s_1}

To derive a dynamic equation for the expansion coefficients $s_1^{\pm 1}$, we project the
dynamic equation (\ref{eq:first_order_field_equation}) for the perturbation
$\delta\phi$ onto the
spherical harmonics $Y_1^{\pm 1}$ [see also eq.\ (\ref{eq:perturbation_ansatz_b})].
Employing the orthonormality relation of the
spherical harmonics and using eqs.\ (\ref{eq:integrals}),
we ultimately obtain
\begin{equation}
  \partial_{t}s_1^{\pm 1}=s_1^{\pm 1}\left[-2b -\left(\frac 3 5 - c\right) \frac M {\sqrt{20\pi}} s_2^0- \kappa\right]+\frac{\xi}{a}\zeta_1^{\pm 1}\;
\label{eq:fluc_rot}
\end{equation}
with noise components $\zeta_l^m$ defined in eq.\ (\ref{eq:fluc_diss_components}).
Due to the nonlinear advection term $M\phi\mathbf{u}|_R$ in eq.\ (\ref{eq:conti_dimless}), 
the coefficients $s_1^{\pm 1}$ couple to $s_2^0$.
The term in square brackets on the right-hand side describes a relaxational dynamics for $s_1^{\pm 1}$.
In particular, for the parameters chosen we find  the swimming droplet to be a pusher. Thus, according 
to eq.\ (\ref{eq:squirmer_parameter}) the coefficient $s_2^0 > 0$ and the term in square brackets is always negative.
On time scales larger than the relaxation time, we can ignore the relaxational dynamics and the time dependence of the  order 
parameter perturbation is solely determined by the thermal noise term, which confirms relation\ (\ref{eq:projected_perturbation}).

Note that in the dynamic equation for $s_l^0$ equivalent to eq.\ (\ref{eq:fluc_rot}),
the advective term $\propto M$ is always positive and triggers for $l=1$ 
and for sufficiently large $M$ the onset of forward propulsion of the droplet 
(see fig.\ \ref{fig:3} and ref. \cite{schmitt2013}).

\end{appendix}
\bibliographystyle{epj_no_urls}
\bibliography{library}

\end{document}